\documentclass[aip,amsmath,amssymb,reprint]{revtex4-1}

\usepackage{graphicx}
\usepackage{dcolumn}
\usepackage{bm}

\usepackage[utf8]{inputenc}
\usepackage[T1]{fontenc}
\usepackage{mathptmx}
\usepackage{etoolbox}

\makeatletter
\def\@email#1#2
	{\endgroup
	\patchcmd{\titleblock@produce}
	{\frontmatter@RRAPformat}
	{\frontmatter@RRAPformat{\produce@RRAP{*#1\href{mailto:#2}{#2}}}\frontmatter@RRAPformat}
	{}{}}
\makeatother

\begin{document}
	
\preprint{AIP/123-QED}

\title{Spatiotemporal Diffusion Metamaterials: Theories and Applications}

\author{Jinrong Liu}
\thanks{These authors contributed equally.}
\address{Department of Physics, State Key Laboratory of Surface Physics, and Key Laboratory of Micro and Nano Photonic Structures (MOE), Fudan University, Shanghai 200438, China}

\author{Liujun Xu}
\thanks{These authors contributed equally.}
\email{jphuang@fudan.edu.cn}
\address{Graduate School of China Academy of Engineering Physics, Beijing 100193, China}

\author{Jiping Huang}
\email{ljxu@gscaep.ac.cn}
\address{Department of Physics, State Key Laboratory of Surface Physics, and Key Laboratory of Micro and Nano Photonic Structures (MOE), Fudan University, Shanghai 200438, China}

\begin{abstract}
Diffusion metamaterials with artificial spatial structures have significant potential in controlling energy and mass transfer. Those static structures may lead to functionality and tunability constraints, impeding the application scope of diffusion metamaterials. Dynamic structures, adding the temporal dimension, have recently provided a new possibility for electric charge and heat diffusion regulation. This perspective introduces the fundamental theories and practical constructions of spatiotemporal diffusion metamaterials for achieving nonreciprocal, topological, or tunable properties. Compared with traditional static design, spatiotemporal modulation is promising to manipulate diffusion processes dynamically, with applications of real-time thermal coding and programming. Existing spatiotemporal diffusion explorations are primarily at macroscopic systems, and we may envision extending these results to microscale and other physical domains like thermal radiation and mass diffusion shortly.
\end{abstract}

\maketitle

\section{Introduction}

Metamaterials have revolutionized the capability to manipulate various wave systems, such as optics~\cite{ChenNM10,CuiJPP24}, acoustics~\cite{CummerNRM16,DongNSR23}, and quantum electrodynamics~\cite{VazquezNC23}. Their characteristic lengths are often the corresponding wavelengths, which are time-independent yet frequency-dependent. Thanks to their elaborate artificial subwavelength structures, varieties of intriguing functionalities, such as cloaking~\cite{LeonhardtScience06,PendryScience06,ChenAPL07-1} and rotating~\cite{ChenAPL07-2}, have been proposed. In contrast to wave processes, diffusion systems have distinctly different governing equations and application scenarios. Diffusionics~\cite{YangRMP24,YangSpringer24} has emerged to manipulate heat and other energy and mass diffusion. Diffusion metamaterials~\cite{XuSpringer23,ZhangNRP23,YangPR21} usually feature time-dependent yet frequency-independent characteristic lengths. Various approaches, such as transformation theory~\cite{FanAPL08}, neutral inclusion/scattering cancellation~\cite{HePRE13}, active control~\cite{NguyenAPL15}, and inverse design~\cite{FujiiAPL18}, have been established to construct spatially distributed parameters, significantly contributing to the flexible control of diffusion fields.

On the other hand, constrained by static structures, conventional metamaterials~\cite{ShenAPL16,XuPRE18,XuNSR23,DaiChaos23} often have only fixed functionalities and almost no tunability, restricting their application scope. Hence, recent advancements have shifted the focus towards spatiotemporal metamaterials~\cite{YineLight22,GaliffiAP22,EnghetaScience23,WaneLight23}. This new class of materials serves as a temporal extension of traditional metamaterials and opens a new avenue to manipulating physical fields with dynamic structures. Exciting developments in electromagnetic and acoustic systems have been achieved, such as nonreciprocal~\cite{CalozPRAP18,NaguluNE20,NassarNRM20,ZhuAPL20,ZhouAPL22} and topological~\cite{NiCR23} transport. However, adding the temporal dimension to previous diffusion metamaterials is particularly challenging due to different governing equations and parameter properties.

\begin{figure}
	\includegraphics[width=0.4\textwidth]{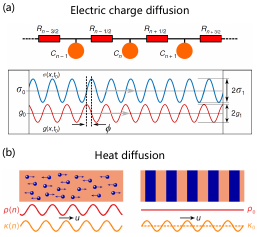}
	\caption{\label{fig1} Concept of spatiotemporal modulation. (a) Electric charge diffusion~\cite{CamachoNC20} with spatiotemporal electric conductivity $\sigma$ and capacity $g$. (b) Heat diffusion~\cite{XuPRL22_2} with spatiotemporal mass density $\rho$ and thermal conductivity $\kappa$.}
\end{figure}

This perspective introduces recent developments in spatiotemporal diffusion metamaterials for remarkable electric charge and heat diffusion control. Briefly, spatiotemporal modulation means that a material parameter has a spatial distribution and dynamically changes in time. Though diffusion processes are governed by similar equations, material parameters may still have different properties. For instance, electric conductivity and capacity can be dynamically modulated in electric charge diffusion [Fig.~\ref{fig1}(a)]. In contrast, heat capacity is often difficult to regulate, and thermal conductivity and mass density are more practical to vary spatiotemporally [Fig.~\ref{fig1}(b)]. Many unexpected functionalities are constructed, such as nonreciprocal diffusion transport~\cite{TorrentPRL18,CamachoNC20,XuPRL22_2}, topological heat transfer~\cite{IJHMTXu21,XuNP22,XuPRL21}, and tunable thermal regulation~\cite{JinAM24,GuoAM22}. We detail their application potential by exploring theoretical underpinnings and practical implementations. We finally provide an outlook on spatiotemporal diffusion metamaterials towards the microscale and other physical domains.

\section{Main Theories}

\subsection{Spatiotemporal Transformation Theory}

Transformation theory is the fundamental of spatiotemporal diffusion metamaterials. From spatial design, traditional transformation theory~\cite{FanAPL08} has been instrumental in projecting the expected space transformation onto material transformation, enabling precise heat flux manipulation~\cite{NarayanaPRL12}. Transformation theory has also been extended to include temperature-dependent material parameters~\cite{LiPRL15,LiPLA16,LeiEPL21} and other heat transfer modes of convection and radiation~\cite{XuPRAP20,XuESEE20}, offering tunable functionalities. One application is thermal coding~\cite{HuPRAP18} pioneered by Hu et al., which switches between thermal cloaking and concentrating to achieve specific spatial modulations, resulting in fixed binary states in unit arrays.

However, traditional transformation theory is based on static parameters, lacking flexibility and efficiency for more advanced applications. The introduction of spatiotemporal transformation theory~\cite{GarciaJP16,YangPRAP22} allows material parameters like thermal conductivity $\kappa$ to be dynamically modulated in both temporal and spatial dimensions, introducing new control freedom. 
The spatiotemporal feature enables the system to Typically it is a quasi-static frequency-independent process that the shell part of the system adapt instantaneously to changing environmental conditions, allowing a single structure to switch between multiple functionalities. For example, spatiotemporal modulation enables coding units to alternate functionalities~\cite{YangPRAP23,LeiIJHMT23}, creating varying coding sequences over time, significantly improving the adaptability and intelligence of information processing systems.

\subsection{Wavelike Diffusion Theory}

Another critical aspect of spatiotemporal diffusion metamaterials lies in wavelike diffusion fields~\cite{MandelisSpringer13}, where the whole system is modulated with frequency-dependent response. For instance, temperature (or concentration) has a periodic spatiotemporal distribution like wave propagation, which can be mathematically represented as \( T= {\rm e}^{{\rm i}(\beta x - \omega t)} + T_0 \), where \( \beta \) and \( \omega \) denote the wave number and frequency. This property can be initiated by an oscillating system or a dynamic convection medium~\cite{JuAM23}, resulting in periodic field changes. Due to the inherent dissipative nature of heat transfer, these systems often experience rapid decay, leading to significant imaginary components in \( \beta \) or \( \omega \).

Homogenization based on plane wave expansion is a primary method for wavelike diffusion fields with spatiotemporal parameter modulation. After homogenization, the traditional Fourier law describing heat conduction should be unexpectedly modified, inducing intriguing nonreciprocal properties. 
For example, the homogenized diffusion parameter changes \(\Delta(\sigma g) =\frac{1}{2}\sigma_1g_1\cos{\phi}\) [Fig.~\ref{fig1}(a)] in rapid oscillations, and the thermal Willis coupling shows unexpected spatiotemporal terms [Fig.~\ref{fig1}(b)].
Nonreciprocity generically indicates that the forward and backward transport is different, which was realized by nonlinear and asymmetric structures~\cite{LiPRL15}. One mechanism of nonreciprocity is the effective advection effect induced by spatiotemporal modulation. This way is robust in both static~\cite{TorrentPRL18,CamachoNC20} and periodic~\cite{XuPRE21} excitation but requires to modulate conductivity and capacity simultaneously~\cite{LiNC22}. Another mechanism of nonreciprocity is the thermal Willis coupling generated by spatiotemporal modulation. This way crucially relies on wavelike diffusion fields but usually needs to modulate only one parameter like conductivity~\cite{XuPRL22,XuPRL22_2,LiPRL22}.

Hamiltonian description~\cite{LiScience19} is another efficient method for wavelike diffusion fields, which reveals non-Hermitian characteristics unique to traditional convection~\cite{LiAM20}. Many intriguing phenomena, such as topological edge states~\cite{XuPRL21,QiAM22,HuAM22,XuNC23} and skin effect~\cite{CaoCP21,CaoCPL22,HuangCPL23}, are proposed through constructing an effective Hamiltonian, opening a new avenue to thermal diffusion behavior.

\section{Practical Structures}

\begin{figure}
    \includegraphics[width=0.4\textwidth]{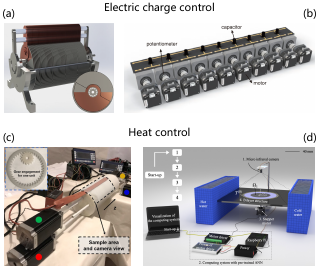}
    \caption{\label{fig2} Practical ways of spatiotemporal modulation. (a) Rotating disks with spatiotemporal capacitors and resistors~\cite{CamachoNC20}. (b) Equivalent circuit with spatiotemporal electric conductivity~\cite{LiPRL22}. (c) Spatiotemporal convection-assisted thermal diffusion~\cite{XuNP22}. (d) Deep learning-assisted heat control~\cite{JinAM24}.}
\end{figure}

Active structures and field-responsive materials are two distinct yet complementary approaches to practically implementing spatiotemporal diffusion metamaterials. Both methods are instrumental in developing materials with dynamic properties that vary across time and space, leading to groundbreaking applications in diffusion systems.

Active structures often include tunable mechanical velocity~\cite{LiOE20,XuPNAS23} or thermoelectric coolers~\cite{HeScience17,GaoCPL23}. For example, the simultaneous modulation of electric conductivity and capacity can be realized by rotating disks [Fig.~\ref{fig2}(a)]. Spatiotemporal thermal conductivity and mass density can be achieved similarly using rotating disks~\cite{LiNC22}. 
The inhomogeneous disks rotate perpendicular to the direction of propagation. The phase difference between the neighboring rotation structures results in the spatial variation, which induces nonreciprocal diffusion.
Another efficient way to achieve spatiotemporal electric conductivity is based on slide rheostats, which are also driven by motors [Fig.~\ref{fig2}(b)]. Since thermal convection~\cite{JuAM23} produces an effective temporal factor, it was arranged in a periodic array to achieve topological thermal transport [Fig.~\ref{fig2}(c)]. Thermal convection can also be combined with a deep learning approach to achieve real-time tunable heat regulation [Fig.~\ref{fig2}(d)]. Besides tunable mechanical velocities, thermoelectric coolers contribute to flexible temperature manipulation by constructing a thermal metasurface~\cite{GuoAM22} with independently controllable units.

Field-dependent materials also stand at the forefront, offering diverse mechanisms for dynamic parameter manipulation in macroscopic and microscopic systems. In macroscopic systems, Strontium Titanate (SrTiO3) cuboids demonstrate temperature-responsive behaviors, facilitating crucial applications like frequency-agile invisibility in transformation optics~\cite{PengPRX17}. Their ability to adapt frequency bands in response to temperature shifts eliminates the need for complex redesigns, streamlining the process and broadening application potential. Meanwhile, in microscopic systems, at extremely low temperatures around 4 K, an experiment demonstrates the Ettingshausen effect~\cite{VolklNaturePhys24}, where a transverse thermoelectric effect develops in the presence of a perpendicular magnetic field, which is applicable for spatiotemporal heat control in the van der Waals semimetals~\cite{AharonSteinbergNature21}.
Moreover, phase-change materials with thermally responsive conductivity are instrumental in maintaining consistent internal temperatures under varying environmental conditions~\cite{ShenPRL16}, a vital feature for energy-efficient thermal systems. Reconfigurable two-phase thermal metamaterials, combining solid-liquid hybrid structures and micromagnetic particle fluids, offer anisotropic thermal conductivity by manipulating radial fluid dynamics~\cite{JinPNAS23} or discrete particle distributions~\cite{XuAMT21}. This temporal versatility is essential in designing intelligent, adaptable devices for various environments.

\section{Distinct Applications}

We introduce the diverse applications of spatiotemporal diffusion metamaterials from three primary domains: nonreciprocal heat transfer, dynamic thermal coding, and intelligent thermal management.

Nonreciprocal heat transfer~\cite{TorrentPRL18} is a pivotal application of spatiotemporal modulation. With the effective advection effect, these materials function adeptly in static~\cite{TorrentPRL18,CamachoNC20} and periodic~\cite{XuPRE21} excitation. Another nonreciprocal advancement for wavelike temperature fields introduces the Willis coupling~\cite{XuPRL22_2}, a notion typically associated with wave systems. A convection-based three-port system can facilitate nonreciprocal heat circulation by a velocity bias~\cite{XuAPL21}. This mechanism is also generalized to deal with steady-state and transient cases by the thermal scattering theory~\cite{JuAM24}. Wavelike diffusion unveils a fresh perspective on directional nonequilibrium energy and mass transfer, offering innovative strategies for manipulating diffusion processes.

Spatiotemporal modulation also advances dynamic thermal coding. In a spatiotemporal thermal binary coding system, coding units equipped with time-dependent metashells experience transitions between cloaking and concentrating, heralding a significant advancement in thermal signal regulation~\cite{YangPRAP23}. 
Furthermore, the integration of microscopic thermoelectric cooling, exemplified by the magneto-thermoelectric effect in nanoscale cryogenic imaging~\cite{VolklNaturePhys24}, showcases critical advancements in device-specific temperature control at cryogenic temperatures.
This capability can be extended to spatiotemporal multiphysics metamaterials~\cite{LeiMTP23}, which simultaneously alternate thermal and electric functionalities, representing a considerable step towards versatile and intelligent field manipulation and multichannel coding even in van der Waals devices.

For thermal management, the temporal dimension further facilitates adaptive structures~\cite{HongAFM20,FangCrystals23} espeically in the radiation systems. In microscale systems, the integration of nonreciprocal thermal photonics~\cite{ZhangPRApplied22}, exploiting near-field photonic thermal diode~\cite{FengAPL21} and time-variant radiative heat transfer~\cite{YuPNAS24}, enables innovative nonreciprocal heat radiation without relying on temperature-dependent optical properties. In macroscopic systems, we build upon the concept of thermal illusion first proposed by Hu et al.~\cite{HuAdvMater18,XiNatComm23}. Through temporal modulation on magnetic polariton resonance~\cite{LiuNanoPhotonics20} or movable structures~\cite{JinResearch23}, these dynamic platforms enable the dynamic functionality of thermal camouflage.
These results offer broad applications for enhancing thermal performance across diverse and changing thermal environments.

\section{outlook}

\begin{figure}
    \includegraphics[width=0.4\textwidth]{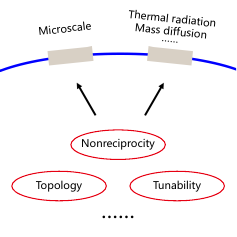}
    \caption{\label{fig3} Outlook of spatiotemporal modulation. Various functionalities like nonreciprocity, topology, and tunability could be extended to the microscale and other physical domains like thermal radiation and mass diffusion.}
\end{figure}

In this perspective, we have delved into the evolving realm of spatiotemporal diffusion metamaterials, a critical subset of diffusion metamaterials, marking a new strategy in manipulating diffusion fields. These innovative materials could pave the way for adaptive architectural solutions and revolutionize diffusion systems by introducing nonreciprocal, topological, and tunable properties. Future explorations (Fig.~\ref{fig3}) might be expected on the microscale and in other physical domains like thermal radiation~\cite{XuPRAP19,LiuNM23,GhanekarACS23,ZhangCPL23} and mass diffusion~\cite{RestrepoAPL17,XuPRE20,ZhangCPL22,ZhangPRAP23}. More potential applications are available, from wearable interactive thermal devices enhancing personal comfort to advanced medical treatments via blood flow regulation. These advancements reflect a significant leap beyond conventional materials, offering dynamic functionalities that include real-time adjustments and complex spatiotemporal transformations. The integration of active structures and field-dependent materials facilitates sophisticated and multifunctional control. In the future, spatiotemporal diffusion metamaterials will stand on the leading edge of reshaping conventional diffusion control, heralding sustainable and efficient solutions for modern society.

\section{acknowledgements}

We are grateful for the revision suggestions put forward by M. Lei. We gratefully acknowledge funding from the National Natural Science Foundation of China (Grants No. 12035004 and No. 12320101004) and the Innovation Program of Shanghai Municipal Education Commission (Grant No. 2023ZKZD06). L.X. acknowledges financial support from the National Natural Science Foundation of China under Grant Nos. 12375040, 12088101, and U2330401.


\begin{thebibliography}{}

\bibitem{ChenNM10} H. Chen, C. T. Chan, and P. Sheng, Nat. Mater. \textbf{9}, 387 (2010).
\bibitem{CuiJPP24} T. J. Cui, et al., J. Phys. Photonics, In Press (2024).


\bibitem{CummerNRM16} S. A. Cummer, J. Christensen, and A. Alù, Nat. Rev. Mater. \textbf{1}, 16001 (2016).
\bibitem{DongNSR23} E. Dong, P. Cao, J. Zhang, S. Zhang, N. X. Fang, and Y. Zhang, Natl. Sci. Rev. \textbf{10}, nwac246 (2023).


\bibitem{VazquezNC23} J. E. V{\'a}zquez-Lozano and I. Liberal, Nat. Commun. \textbf{14}, 4606 (2023).


\bibitem{LeonhardtScience06} U. Leonhardt, Science \textbf{312}, 1777 (2006).
\bibitem{PendryScience06} J. B. Pendry, D. Schurig, and D. R. Smith, Science \textbf{312}, 1780 (2006).
\bibitem{ChenAPL07-1} H. Chen and C. T. Chan, Appl. Phys. Lett. \textbf{91}, 183518 (2007).


\bibitem{ChenAPL07-2} H. Chen and C. T. Chan, Appl. Phys. Lett. \textbf{90}, 241105 (2007).


\bibitem{YangRMP24} F. B. Yang, Z. R. Zhang, L. J. Xu, Z. F. Liu, P. Jin, P. F. Zhuang, M. Lei, J. R. Liu, J.-H. Jiang, X. P. Ouyang, F. Marchesoni, and J. P. Huang, Rev. Mod. Phys. \textbf{96}, 015002 (2024).
\bibitem{YangSpringer24} F. B. Yang and J. P. Huang, \textit{Diffusionics: Diffusion Process Controlled by Diffusion Metamaterials} (Springer, Singapore, 2024).


\bibitem{XuSpringer23} L. J. Xu and J. P. Huang, \textit{Transformation Thermotics and Extended Theories: Inside and Outside Metamaterials} (Springer, Singapore, 2023).
\bibitem{ZhangNRP23} Z. R. Zhang, L. J. Xu, T. Qu, M. Lei, Z.-K. Lin, X. P. Ouyang, J.-H. Jiang, and J. P. Huang, Nat. Rev. Phys. \textbf{5}, 218 (2023).
\bibitem{YangPR21} S. Yang, J. Wang, G. Dai, F. Yang, and J. Huang, Phys. Rep. \textbf{908}, 1 (2021).


\bibitem{FanAPL08} C. Z. Fan, Y. Gao, and J. P. Huang, Appl. Phys. Lett. \textbf{92}, 251907 (2008).
\bibitem{HePRE13} X. He and L. Wu, Phys. Rev. E \textbf{88}, 033201 (2013).
\bibitem{NguyenAPL15} D. M. Nguyen, H. Xu, Y. Zhang, and B. Zhang, Appl. Phys. Lett. \textbf{107}, 121901 (2015).
\bibitem{FujiiAPL18} G. Fujii, Y. Akimoto, and M. Takahashi, Appl. Phys. Lett. \textbf{112}, 061108 (2018).


\bibitem{ShenAPL16} X. Y. Shen, C. R. Jiang, Y. Li, and J. P. Huang, Appl. Phys. Lett. \textbf{109}, 201906 (2016).
\bibitem{XuPRE18} L. J. Xu, S. Yang, and J. P. Huang, Phys. Rev. E \textbf{98}, 052128 (2018).
\bibitem{XuNSR23} L. J. Xu, J. R. Liu, P. Jin, G. Q. Xu, J. X. Li, X. P. Ouyang, Y. Li, C.-W. Qiu, and J. P. Huang, Natl. Sci. Rev. \textbf{10}, nwac159 (2023).
\bibitem{DaiChaos23} G. L. Dai, F. B. Yang, J. Wang, L. J. Xu, and J. P. Huang, Chaos, Solitons \& Fractals \textbf{174}, 113849 (2023).


\bibitem{YineLight22} S. Yin, E. Galiffi, and A. Alù, eLight \textbf{2}, 8 (2022).
\bibitem{GaliffiAP22} E. Galiffi, R. Tirole, S. Yin, H. Li, S. Vezzoli, P. A. Huidobro, M. G. Silveirinha, R. Sapienza, A. Alù, and J. B. Pendry, Adv. Photonics \text{4}, 014002 (2022).
\bibitem{EnghetaScience23} N. Engheta, Science \textbf{379}, 1190 (2023).
\bibitem{WaneLight23} C. Wan, A. Chong, and Q. Zhan, eLight \textbf{3}, 11 (2023).


\bibitem{CalozPRAP18} C. Caloz, A. Alù, S. Tretyakov, D. Sounas, K. Achouri, and Z.-L. Deck-Léger, Phys. Rev. Appl. \textbf{10}, 047001 (2018).
\bibitem{NaguluNE20} A. Nagulu, N. Reiskarimian, and H. Krishnaswamy, Nat. Electron. \textbf{3}, 241 (2020).
\bibitem{NassarNRM20} H. Nassar, B. Yousefzadeh, R. Fleury, M. Ruzzene, A. Alù, C. Daraio, A. N. Norris, G. Huang, and M. R. Haberman, Nat. Rev. Mater. \textbf{5}, 667 (2020).
\bibitem{ZhuAPL20} X. Zhu, J. Li, C. Shen, X. Peng, A. Song, L. Li, and S. A. Cummer, Appl. Phys. Lett. \textbf{116}, 034101 (2020).
\bibitem{ZhouAPL22} H. Zhou and A. Baz, Appl. Phys. Lett. \textbf{121}, 061701 (2022).


\bibitem{NiCR23} X. Ni, S. Yves, A. Krasnok, and A. Alù, Chem. Rev. \textbf{123}, 7585 (2023).


\bibitem{TorrentPRL18} D. Torrent, O. Poncelet, and J. C. Batsale, Phys. Rev. Lett. \textbf{120}, 125501 (2018).
\bibitem{CamachoNC20} M. Camacho, B. Edwards, and N. Engheta, Nat. Commun. \textbf{11}, 3733 (2020).
\bibitem{XuPRL22_2} L. J. Xu, G. Q. Xu, J. X. Li, Y. Li, J. P. Huang, and C.-W. Qiu, Phys. Rev. Lett. \textbf{129}, 155901 (2022).


\bibitem{IJHMTXu21} L. J. Xu, J. Wang, G. L. Dai, S. Yang, F. B. Yang, G. Wang, and J. P. Huang, Int. J. Heat Mass Transf. \textbf{165}, 120659 (2021).
\bibitem{XuNP22} G. Xu, Y. Yang, X. Zhou, H. Chen, A. Alù, and C.-W. Qiu, Nat. Phys. \textbf{18}, 450 (2022).
\bibitem{XuPRL21} G. Xu, Y. Li, W. Li, S. Fan, and C.-W. Qiu, Phys. Rev. Lett. \textbf{127}, 105901 (2021).


\bibitem{JinAM24} P. Jin, L. J. Xu, G. Q. Xu, J. X. Li, C.-W. Qiu, and J. P. Huang, Adv. Mater. \textbf{36}, 2305791 (2024).
\bibitem{GuoAM22} J. Guo, G. Xu, D. Tian, Z. Qu, and C.-W. Qiu, Adv. Mater. \textbf{34}, 2201093 (2022).



\bibitem{NarayanaPRL12} S. Narayana and Y. Sato, Phys. Rev. Lett. \textbf{108}, 214303 (2012).


\bibitem{LiPRL15} Y. Li, X. Y. Shen, Z. H. Wu, J. Y. Huang, Y. X. Chen, Y. S. Ni, and J. P. Huang, Phys. Rev. Lett. \textbf{115}, 195503 (2015). 
\bibitem{LiPLA16} Y. Li, X. Y. Shen, J. P. Huang, and Y. S. Ni, Phys. Lett. A \textbf{380}, 1641 (2016).
\bibitem{LeiEPL21} M. Lei, J. Wang, G. L. Dai, P. Tan, and J. P. Huang, EPL \textbf{135}, 54003 (2021).


\bibitem{XuPRAP20} L. J. Xu, G. L. Dai, and J. P. Huang, Phys. Rev. Appl. \textbf{13}, 024063 (2020).
\bibitem{XuESEE20} L. J. Xu, S. Yang, G. L. Dai, and J. P. Huang, ES Energy Environ. \textbf{7}, 65 (2020).


\bibitem{HuPRAP18} R. Hu, S. Huang, M. Wang, L. Zhou, X. Peng, and X. Luo, Phys. Rev. Appl. \textbf{10}, 054032 (2018).


\bibitem{GarciaJP16} C. García-Meca and C. Barceló, J. Opt. \textbf{18}, 044026 (2016).
\bibitem{YangPRAP22} F. B. Yang, L. J. Xu, J. Wang, and J. P. Huang, Phys. Rev. Appl. \textbf{18}, 034080 (2022).


\bibitem{YangPRAP23} F. B. Yang, P. Jin, M. Lei, G. L. Dai, J. Wang, and J. P. Huang, Phys. Rev. Appl. \textbf{19}, 054096 (2023).
\bibitem{LeiIJHMT23} M. Lei, C. R. Jiang, F. B. Yang, J. Wang, and J. P. Huang, Int. J. Heat Mass Transf. \textbf{207}, 124033 (2023).



\bibitem{MandelisSpringer13} A. Mandelis, \textit{Diffusion-Wave Fields: Mathematical Methods and Green Functions} (Springer Science \& Business Media, 2013).

\bibitem{JuAM23} R. Ju, G. Xu, L. Xu, M. Qi, D. Wang, P.-C. Cao, R. Xi, Y. Shou, H. Chen, C.-W. Qiu, and Y. Li, Adv. Mater. \textbf{35}, 2209123 (2023).


\bibitem{XuPRE21} L. J. Xu, J. P. Huang, and X. P. Ouyang, Phys. Rev. E \textbf{103}, 032128 (2021).
\bibitem{LiNC22} J. Li, Y. Li, P.-C. Cao, M. Qi, X. Zheng, Y.-G. Peng, B. Li, X.-F. Zhu, A. Alù, H. Chen, and C.-W. Qiu, Nat. Commun. \textbf{13}, 167 (2022).


\bibitem{XuPRL22} L. J. Xu, G. Q. Xu, J. P. Huang, and C.-W. Qiu, Phys. Rev. Lett. \textbf{128}, 145901 (2022).
\bibitem{LiPRL22} J. Li, Z. Zhang, G. Xu, H. Sun, L. Dai, T. Li, and C.-W. Qiu, Phys. Rev. Lett. \textbf{129}, 256601 (2022).


\bibitem{LiScience19} Y. Li, Y.-G. Peng, L. Han, M.-A. Miri, W. Li, M. Xiao, X.-F. Zhu, J. Zhao, A. Alù, S. Fan, and C.-W. Qiu, Science \textbf{364}, 170 (2019).


\bibitem{LiAM20} J. Li, Y. Li, P.-C. Cao, T. Yang, X.-F. Zhu, W. Wang, and C.-W. Qiu, Adv. Mater. \textbf{32}, 2003823 (2020).


\bibitem{QiAM22} M. Qi, D. Wang, P.-C. Cao, X.-F. Zhu, C.-W. Qiu, H. Chen, and Y. Li, Adv. Mater. \textbf{34}, 2202241 (2022).
\bibitem{HuAM22} H. Hu, S. Han, Y. Yang, D. J. Liu, H. Xue, G.-G. Liu, Z. Cheng, Q. J. Wang, S. Zhang, B. Zhang, and Y. Luo, Adv. Mater. \textbf{34}, 2202257 (2022).
\bibitem{XuNC23} G. Xu, X. Zhou, S. Yang, J. Wu, and C.-W. Qiu, Nat. Commun. \textbf{14}, 3252 (2023).


\bibitem{CaoCP21} P.-C. Cao, Y. Li, Y.-G. Peng, M. Qi, W.-X. Huang, P.-Q. Li, and X.-F. Zhu, Commun. Phys. \textbf{4}, 230 (2021).
\bibitem{CaoCPL22} P.-C. Cao, Y.-G. Peng, Y. Li, and X.-F. Zhu, Chin. Phys. Lett. \textbf{39}, 057801 (2022).
\bibitem{HuangCPL23} Q.-K.-L. Huang, Y.-K. Liu, P.-C. Cao, X.-F. Zhu, and Y. Li, Chin. Phys. Lett. \textbf{40}, 106601 (2023).



\bibitem{LiOE20} J. Li, Y. Li, W. Wang, L. Li, and C.-W. Qiu, Opt. Express \textbf{28}, 25894 (2020).
\bibitem{XuPNAS23} L. J. Xu, J. R. Liu, G. Q. Xu, J. P. Huang, and C.-W. Qiu, Proc. Natl. Acad. Sci. U.S.A. \textbf{120}, e2305755120 (2023).


\bibitem{HeScience17} J. He and T. M. Tritt, Science \textbf{357}, eaak9997 (2017).
\bibitem{GaoCPL23} J.-Z. Gao, X. Liu, J.-H. Wang, and J.-Z. He, Chin. Phys. Lett. \textbf{40}, 117301 (2023).


\bibitem{PengPRX17} R. Peng, Z. Xiao, Q. Zhao, F. Zhang, Y. Meng, B. Li, J. Zhou, Y. Fan, P. Zhang, N.-H. Shen, et al., Phys. Rev. X \textbf{7}, 011033 (2017).
\bibitem{VolklNaturePhys24} T. Völkl, A. Aharon-Steinberg, T. Holder, E. Alpern, N. Banu, A. K. Pariari, Y. Myasoedov, M. E. Huber, M. Hücker, E. Zeldov, Nature Phys. 1 (2024).
\bibitem{AharonSteinbergNature21} A. Aharon-Steinberg, A. Marguerite, D. J. Perello, K. Bagani, T. Holder, Y. Myasoedov, L. S. Levitov, A. K. Geim, E. Zeldov, Nature \textbf{593}, 528 (2021).
\bibitem{ShenPRL16} X. Y. Shen, Y. Li, C. R. Jiang, and J. P. Huang, Phys. Rev. Lett. \textbf{117}, 055501 (2016).

\bibitem{JinPNAS23} P. Jin, J. R. Liu, L. J. Xu, J. Wang, X. P. Ouyang, J.-H. Jiang, and J. P. Huang, Proc. Natl. Acad. Sci. U.S.A. \textbf{120}, e2217068120 (2023).

\bibitem{XuAMT21} Z. Xu, L. Li, X. Chang, Y. Zhao, and W. Wang, Appl. Mater. Today \textbf{22}, 100911 (2021).


\bibitem{XuAPL21} L. J. Xu, J. P. Huang, and X. P. Ouyang, Appl. Phys. Lett. \textbf{118}, 221902 (2021).

\bibitem{JuAM24} R. Ju, P. C. Cao, D. Wang, M. Qi, L. J. Xu, S. Yang, C.-W. Qiu, H. Chen, and Y. Li, Adv. Mater. \textbf{36}, 2309835 (2024).


\bibitem{LeiMTP23} M. Lei, L. J. Xu, and J. P. Huang, Mater. Today Phys. \textbf{34}, 101057 (2023).


\bibitem{HongAFM20} S. Hong, S. Shin, and R. Chen, Adv. Funct. Mater. \textbf{30}, 1909788 (2020).
\bibitem{FangCrystals23} J. Fang, S. Shi, K. Sun, C. Di, Y. Lin, Y. Zhu, S. Zhang, and Y. Shi, Crystals \textbf{13}, 996 (2023).
\bibitem{ZhangPRApplied22} Z. Zhang, L. Zhu, Phys. Rev. Applied \textbf{18}, 027001 (2022).
\bibitem{FengAPL21} Z. Chen, S. Yu, C. Yuan, X. Luo, R. Hu, Int. J. Heat Mass Transfer \textbf{222}, 125202 (2024).
\bibitem{YuPNAS24} R. Yu, S. Fan, Proc. Natl. Acad. Sci. U.S.A.\textbf{121}, e2401514121 (2024).
\bibitem{HuAdvMater18} R. Hu, S. Zhou, Y. Li, D.-Y. Lei, X. Luo, C.-W. Qiu, Adv. Mater. \textbf{30}, 1707237 (2018).
\bibitem{XiNatComm23} W. Xi, Y.-J. Lee, S. Yu, Z. Chen, J. Shiomi, S.-K. Kim, R. Hu, Nat. Commun. \textbf{14}, 4694 (2023).
\bibitem{LiuNanoPhotonics20} Y. Liu, J. Song, W. Zhao, X. Ren, Q. Cheng, X. Luo, N. X. Fang, R. Hu, Nanophotonics \textbf{9}, 855-863 (2020).
\bibitem{JinResearch23} P. Jin, J. Liu, F. Yang, F. Marchesoni, J.-H. Jiang, J. Huang, Research \textbf{6}, 0222 (2023).

\bibitem{XuPRAP19} L. J. Xu and J. P. Huang, Phys. Rev. Appl. \textbf{12}, 044048 (2019).
\bibitem{LiuNM23} M. Liu, S. Xia, W. Wan, J. Qin, H. Li, C. Zhao, L. Bi, and C.-W. Qiu, Nat. Mater. \textbf{22}, 1196 (2023).
\bibitem{GhanekarACS23} A. Ghanekar, J. Wang, C. Guo, S. Fan, and M. L. Povinelli, ACS Photonics \textbf{10}, 170 (2023).
\bibitem{ZhangCPL23} C. X. Zhang, T. J. Li, L. J. Xu, and J. P. Huang, Chin. Phys. Lett. \textbf{40}, 054401 (2023).


\bibitem{RestrepoAPL17} J. M. Restrepo-Flórez and M. Maldovan, Appl. Phys. Lett. \textbf{111}, 071903 (2017).
\bibitem{XuPRE20} L. J. Xu, G. L. Dai, G. Wang, and J. P. Huang, Phys. Rev. E \textbf{102}, 032140 (2020) 
\bibitem{ZhangCPL22} Z. R. Zhang and J. P. Huang, Chin. Phys. Lett. \textbf{39}, 075201 (2022).
\bibitem{ZhangPRAP23} Z. R. Zhang, F. B. Yang, and J. P. Huang, Phys. Rev. Appl. \textbf{19}, 024009 (2023). 

\end{thebibliography}
\end{document}